# SOLAR NEUTRINO VARIABILITY AND ITS IMPLICATIONS FOR SOLAR PHYSICS AND NEUTRINO PHYSICS

P.A. Sturrock

Center for Space Science Astrophysics, Varian 302, Stanford University,

Stanford, CA 94305-4060, U.S.A.

## ABSTRACT

Recent coordinated power-spectrum analyses of radiochemical solar neutrino data and the solar irradiance have revealed a highly significant, high-Q common modulation at 11.85 yr$^{-1}$. Since the stability of this frequency points to an explanation in terms of rotation, this result may be attributable to non-spherically-symmetric nuclear burning in a solar core with sidereal rotation frequency 12.85 yr$^{-1}$. The variability of the amplitude (on a timescale of years) suggests that the relevant nuclear burning is variable as well as asymmetric. Recent analysis of Super-Kamiokande solar neutrino data has revealed r-mode-type modulations with frequencies corresponding to a region with sidereal rotation frequency 13.97 yr$^{-1}$. If this modulation is attributed to the RSFP (Resonant Spin Flavor Precession) process, it provides a measurement of the rotation rate deep in the radiative zone. These two results suggest that the core rotates significantly more slowly than the radiative zone. If one accepts an upper limit of 7 MG for the Sun's internal magnetic field, an RSFP interpretation of the Super-Kamiokande results leads to a lower limit of $10^{-12}$ Bohr magnetons for the neutrino transition magnetic moment.

*Subject headings:* neutrinos - Sun:interior - Sun:rotation

# 1. INTRODUCTION

Evidence for variability of the solar neutrino flux - which was studied by Davis (1996), the pioneer in this field - continues to attract attention. Early studies involved tests for correlations with other solar variables, such as sunspot number (Bahcall & Press 1991), magnetic field (Oakley et al. 1994), coronal indexes (Massetti et al. 1997), p-mode measurements (Delache et al. 1993), and solar-wind speed (McNutt 1995). Such a correlation would have provided significant evidence for variability of the solar neutrino flux, but none of the correlations has been generally considered to be sufficiently persuasive.

Our approach has been primarily to apply power spectrum analysis - see, for instance, Sturrock et al. (1997), applied to Homestake data; Sturrock, Caldwell & Scargle (2006) and Sturrock (2008a), applied to GALLEX-GNO data; Sturrock & Scargle (2006), applied to Super-Kamiokande data; and Sturrock (2006), applied to SNO data. Similar analyses have been carried out by several other authors (Haubold & Gerth 1990; Gavryusev et al. 1991; Vasiliev & Ogurtsov 1995; Rivin & Obridko 1997; Kolomeets et al. 1998; Shirai 2004; Nakahata et al. 2005; and Ranucci 2006). These results also have not been generally accepted as providing convincing evidence of variability.

However, two recent analyses have provided surprising new information concerning solar neutrinos. The purpose of this article is to explore the implications of these new results for solar physics and for neutrino physics.

One of the new results, summarized in the Section 2, emerges from a combined power spectrum analysis of radiochemical solar neutrino data (from the Homestake and GALLEX experiments) and irradiance data (from the ACRIM experiment) (Sturrock 2008a). We find that neutrino and irradiance measurements both show strong evidence for a high-Q, periodic signal with frequency 11.85 yr$^{-1}$. Such a stable oscillation strongly suggests that this signal is due to rotation. The corresponding sidereal rotation frequency is 12.85 yr$^{-1}$ (407 nHz). However, this frequency does not conform to the known rotation profiles of the convection zone and the radiative zone (Schou et al., 1998; Garcia et al., 2008), implying that this signal has its origin in rotation of the solar core. This interpretation requires, of course, that the core is not axially symmetric with respect to its rotation axis.

The other recent development, summarized in Section 3, is that power-spectrum analysis of Super-Kamiokande data yields persuasive evidence for temporal variation associated with a group of r-mode oscillations, which may be understood in terms of the RSFP process deep in the radiative zone (Sturrock 2008b). "Resonant spin flavor precession" is a process by which an electron neutrino traversing a magnetic field may be converted into a neutrino of a different flavor (muon or tau), provided the neutrino has a "transition magnetic moment" (Akhmedov 1988; Lim & Marciano 1988).

We discuss the possible implications of these results for solar physics and neutrino physics in Section 4.

## 2. JOINT ANALYSIS OF RADIOCHEMICAL AND IRRADIANCE DATA

We have recently compared power spectra derived from Homestake and GALLEX data and from the corresponding intervals of the ACRIM irradiance data (Sturrock 2008a). We examine the entire GALLEX interval, 1991.367 to 1997.060, but we examine Homestake data from the beginning, 1970.281, only until run 115 that ends at 1991.265, to avoid overlap with the GALLEX experiment. We analyze the radiochemical measurements by the likelihood method (Sturrock et al. 1997). In analyzing the irradiance measurements, we first subtract 600-day running means to remove the solar-cycle modulation, and then use the Lomb-Scargle method of power spectrum analysis (Lomb 1976; Scargle 1982).

Figure 1 shows sections of these four power spectra for the frequency range 10 - 15 yr$^{-1}$, which covers the expected range of synodic rotation frequency for the solar interior (Schou et al. 1998). Each of the four datasets contains significant modulation at or very close to 11.85 yr$^{-1}$. We can conveniently search for a common feature in these power spectra by forming the joint power statistic (JPS). This is a measure of correlation given by a function of the product of power spectra that has the property that, if each power spectrum has an exponential distribution [such that the probability of getting the power $S$ or more is $e^{-S}$], the JPS also has an exponential distribution (Sturrock et al. 2005). For four power measurements $S_1,…,S_4$, this statistic is given to good approximation by

$$J = \frac{3.88\, Y^2}{1.27 + Y}, \qquad (1)$$

where

$$Y = (S_1 * S_2 * S_3 * S_4)^{1/4}. \qquad (2)$$

This quantity is shown as a function of frequency in Figure 2. We find an extraordinarily strong peak, with value J = 40.87, at $\nu = 11.85 \, \text{yr}^{-1}$.

Each of the procedures of power-spectrum analysis would yield an exponential distribution if measurements were independent and drawn from a normal distribution. In that case, the probability of finding a peak with value 40.87 or more within the frequency band 10 - 15 yr$^{-1}$, which is found to contain 66 separate peaks, would be given by $66 \times \exp(-40.87)$, i.e. 1.2 10$^{-16}$. In order to obtain a more robust significance estimate, we would need to take into account the known properties of irradiance measurements, such as their relationship to sunspot and facula measurements (Froehlich & Lean 2004). Variations in irradiance measurements due to sunspots and faculae are much bigger than those due to the modulation at 11.85 yr$^{-1}$, but they do not have a long-term, sharply defined frequency, and do not contribute to that periodicity.

However, it seems reasonable to accept the fact that modulation of the irradiance measurements at $\nu = 11.85 \, \text{yr}^{-1}$ is highly significant, and then estimate the probability that modulation at the same frequency will show up also in the radiochemical neutrino measurements. We may make a robust estimate of this probability by carrying out the shuffle test (Bahcall et al. 1987), in which one randomly re-assigns measurements to times. For each shuffle, we re-calculate the JPS formed from the two (shuffled) sets of neutrino measurements and the two (unshuffled) sets of irradiance measurements. For only 20 simulations out of 100,000 do we find a value of the JPS that is as large as or larger than the value (40.87) derived from the actual data. Hence the case for rotational modulation of both Homestake and GALLEX data at a frequency found with high significance in ACRIM irradiance data may be accepted at the 99.98% confidence level.

## 3. SUPER-KAMIOKANDE MODULATION

We now turn to evidence that some of the modulation of solar neutrinos is due to the RSFP process (Akhmedov 1998; Lim & Marciano 1988). Such modulation is expected to have a depth of modulation (the ratio of the amplitude to the mean value) of a few percent (Chauhan et al. 2005; Balantekin & Volpe 2005). [Modulation found in Homestake (Sturrock et al. 1997) and GALLEX neutrino data (Sturrock 2008a,c) is much larger and is presumably due to some process other than RSFP.] We have recently (Sturrock 2008b) identified modulation in the Super-Kamiokande data (Fukuda et al. 2001, 2002; Fukuda 2003) that has a depth of modulation of only a few percent, consistent with what is expected from the RSFP process. This modulation has been detected in r-modes, a class of waves that can propagate in a rotating, uniform, fluid sphere (Papaloizou & Pringle 1978; Provost et al. 1981; Saio 1982).

If the sphere rotates with frequency $\nu_R$, r-modes are retrograde with respect to the rotating fluid, with frequencies

$$\nu(l,m,S) = \frac{2m\nu_R}{l(l+1)}, \qquad (4)$$

where $l$ and $m$ are two of the three spherical-harmonic indices (the frequency is approximately independent of $n$). The index $l$ takes the values $l = 2, 3,...$, and $m$ takes the values $m = 1,...,l$. We use the symbol "S" to denote that these are the frequencies that would be detected by an observer rotating with the Sun. As seen from Earth, and for waves rotating in the plane of the ecliptic, the frequencies (measured in cycles per year) are given by

$$\nu(l,m,E) = m(\nu_R - 1) - \frac{2m\nu_R}{l(l+1)}. \quad (5)$$

We have found that the following five r-modes, with frequencies as given in Equation (5), show up in the Super-Kamiokande power spectrum: $m = 1, l = 2, 3, 4, 5,$ and $6$ (Sturrock 2008b). In Figure 3, we show the JPS formed from these five frequencies as a function of the (sidereal) rotation frequency. This is given, to good approximation, by

$$J = \frac{4.9\,Y^2}{1.6 + Y}, \quad (6)$$

where

$$Y = (S_1 * .. * S_5)^{1/5}. \quad (7)$$

The biggest peak occurs at 13.97 yr$^{-1}$, with value $J_M = 11.48$. We have found, by Monte Carlo simulations, that this value or more of the joint power statistic would be found by chance, over the radiative-zone-convection-zone rotational frequency band 13 to 15 yr$^{-1}$, with probability only 0.003.

## 4. DISCUSSION

Since the sidereal rotation rate (12.85 yr$^{-1}$ or 407 nHz) inferred in Section 2 is lower than the rotation rate of the radiative zone (mean value 13.9 yr$^{-1}$ or 440 nHz; Garcia et al. 2008) and much lower than that of an equatorial section of the convection zone (and bearing in mind that neutrinos originate in the core), it seems reasonable to attribute the common modulation of the radiochemical neutrinos and the irradiance to rotation of the solar core. This inference is consistent with the theoretical proposal that gravity waves may transfer angular momentum from the core to the outer layers of the Sun, leading to a slowly rotating core (Talon et al. 2002; Charbonnel & Talon 2005).

Concerning the mechanism of this rotational periodicity of the neutrino flux, we first consider the possibility that it may be due to the RSFP process. In terms of the known neutrino flavors, this process can occur inside the solar core for MeV energies, but not for the sub-MeV energies detected by GALLEX (Balantekin & Volpe, 2005). In terms of a possible sterile neutrino, the resonance could occur in the convection zone (Chauhan et al. 2005), but this possibility is unacceptable since the inferred sidereal frequency (12.85 $yr^{-1}$) is too low to be compatible with the rotation profile of an equatorial section of the convection zone (Schou et al. 1998).

It is therefore necessary to look for an alternative interpretation. It seems possible that the modulation at 11.85 $yr^{-1}$ (which has a larger depth of modulation than one would expect from the RSFP process) may be due to nuclear burning that is localized in heliographic longitude. This process may be localized in radius, in which case it would influence certain production processes more than others, and therefore be more evident to one experiment than another. (See, for instance, Bahcall 1989, Fig. 6.1, p. 147.) It is also possible that the nuclear burning is time dependent (see, for instance, Sturrock, Caldwell, & Scargle 2006), with the result that it might show up in data from one experiment but not from another if there is no overlap in the intervals of operation.

This interpretation may also explain the high-Q periodicity found in the irradiance measurements, since asymmetric nuclear burning may lead to a corresponding rotating asymmetry in the temperature of the photosphere. This would be a quite different effect than the modulation (mainly at the solar cycle frequency, and of much larger amplitude) in the irradiance due to sunspots and faculae (Froehlich & Lean 2004), as we have confirmed by a power-spectrum analysis of the sunspot number that shows no notable feature at or near 11.85 yr$^{-1}$.

Turning now to Section 3, we first note that a sidereal rotation frequency of 13.97 yr$^{-1}$ (443 nHz) is compatible with rotation in the radiative zone (Garcia et al. 2008). The few-percent depth of modulation of the Super-Kamiokande flux measurements is consistent with what is to be expected from the RSFP process which – to be compatible with RSFP theory (Balantekin & Volpe 2005) – would need to be occurring just above the outer radius of the core (0.2 $R_{solar}$).

To conform to RSFP calculations (Balantekin & Volpe 2005), we need $\mu B \approx 10^{-5} \mu_B$ Gauss, where $\mu_B$ is the Bohr magneton. However, Friedland & Gruzinov (2006) argue that measurements of the solar oblateness yield an upper bound of 7 MG on the strength of the Sun's internal magnetic field (assumed to be toroidal). This upper limit would lead to a lower limit of $1.4\,10^{-12} \mu_B$ for the neutrino transition magnetic moment. [According to Pulido (private communication), modulation of the Super-Kamiokande flux measurements is not compatible with the sterile-neutrino model (Chauhan et al. 2005).]

The lowest upper-bound on the neutrino transition magnetic moment comes from astrophysical considerations (Raffelt 1990), which lead to the limit $\mu \leq 2.5\,10^{-12} \mu_B$. Hence it seems likely that the magnetic moment is approximately in the range $\mu/\mu_B \approx (1 - 2.5) \times 10^{-12}$ (but note that the lower limit on the magnetic moment scales inversely with the assumed magnetic field strength deep in the radiative zone).

These new results raise many questions. It will be important to compare the periodicity found in Section 2 with information concerning solar rotation within and near the solar core derived from helioseismology, when such information becomes available. If it is found that there is a sharp transition from a slowly rotating core to a faster rotating radiative zone, this could imply that there is a second (inner) tachocline, which could in principle exhibit dynamo action similar to that in the known (outer) tachocline and therefore be a source of magnetic field, and may be related to the excitation of the r-modes detected in Super-Kamiokande data. For both solar physics and neutrino physics, it also seems important to determine – or to put tighter constraints on – the magnetic field strength in and near the solar core.

I wish to thank the GALLEX, GNO, Homestake and Super-Kamiokande consortia for sharing their solar neutrino data, Claus Froehlich and Judith Lean for making available irradiance data, Baha Balantekin, David Caldwell, Joao Pulido and Steven Yellin for helpful discussions concerning neutrino physics, and Sebastien Coudivat, Alexander Kosovichev, Jeffrey Scargle, Guenther Walther and Michael Wheatland for helpful discussions concerning solar physics and statistics. I also thank an anonymous referee whose comments and suggestions have led, I believe, to a significant improvement in this article. This work was supported by NSF Grant AST-0607572.

FIGURES

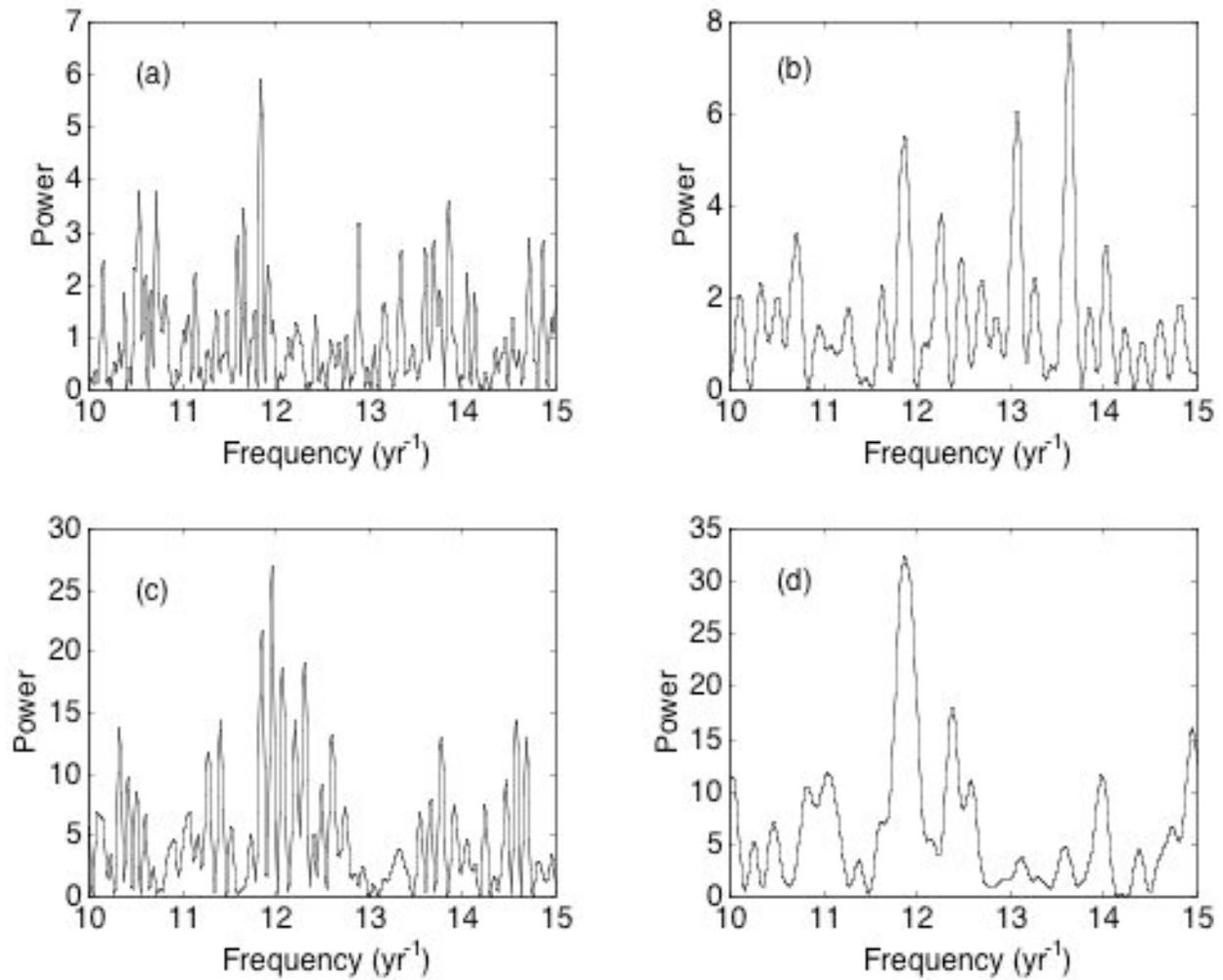

Figure 1. Power spectra formed from (a) Homestake data; (b) GALLEX data; (c) ACRIM data for the Homestake interval; and (d) ACRIM data for the GALLEX interval.

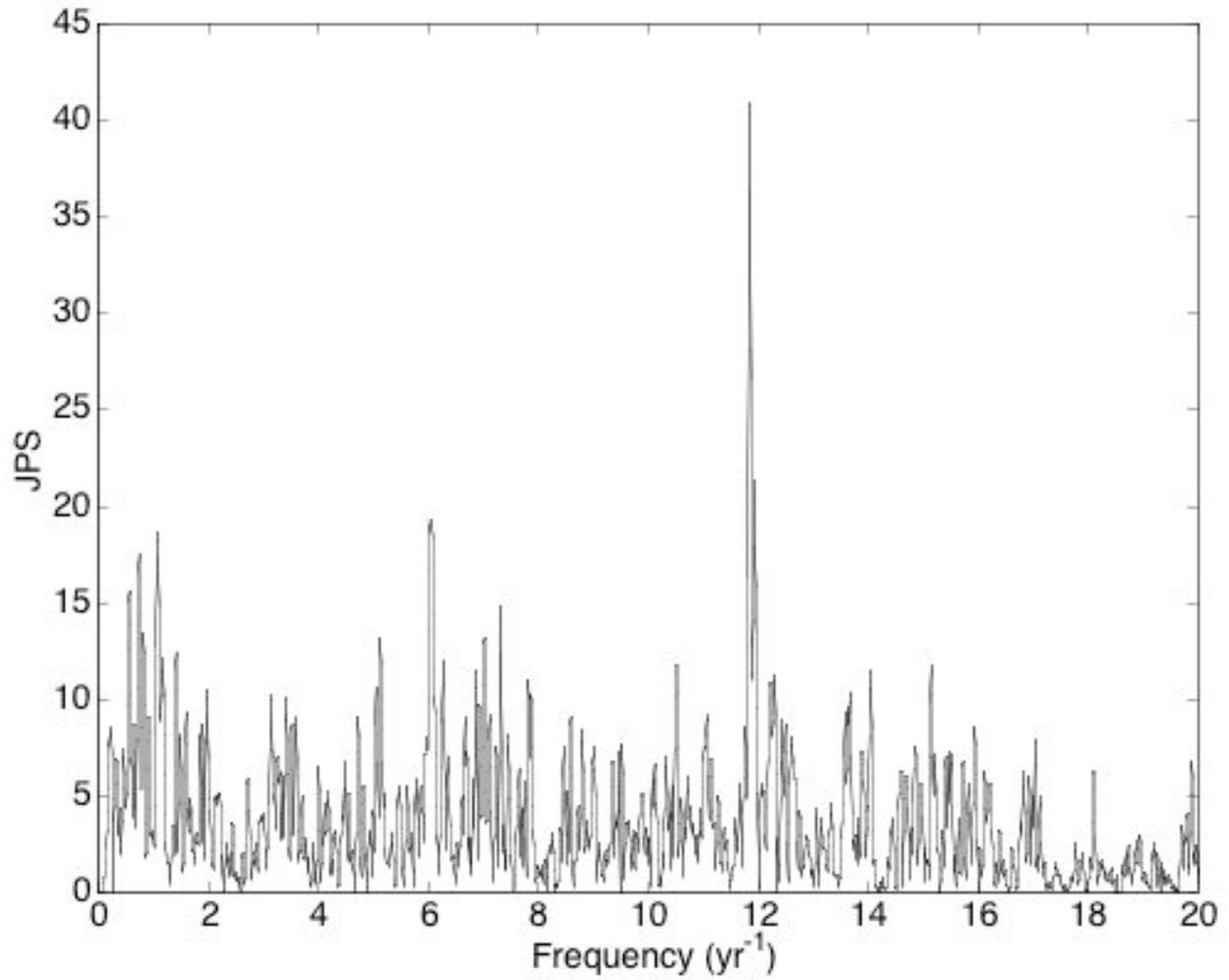

Figure 2. Joint power statistic formed from the four power spectra shown in Figure 1.

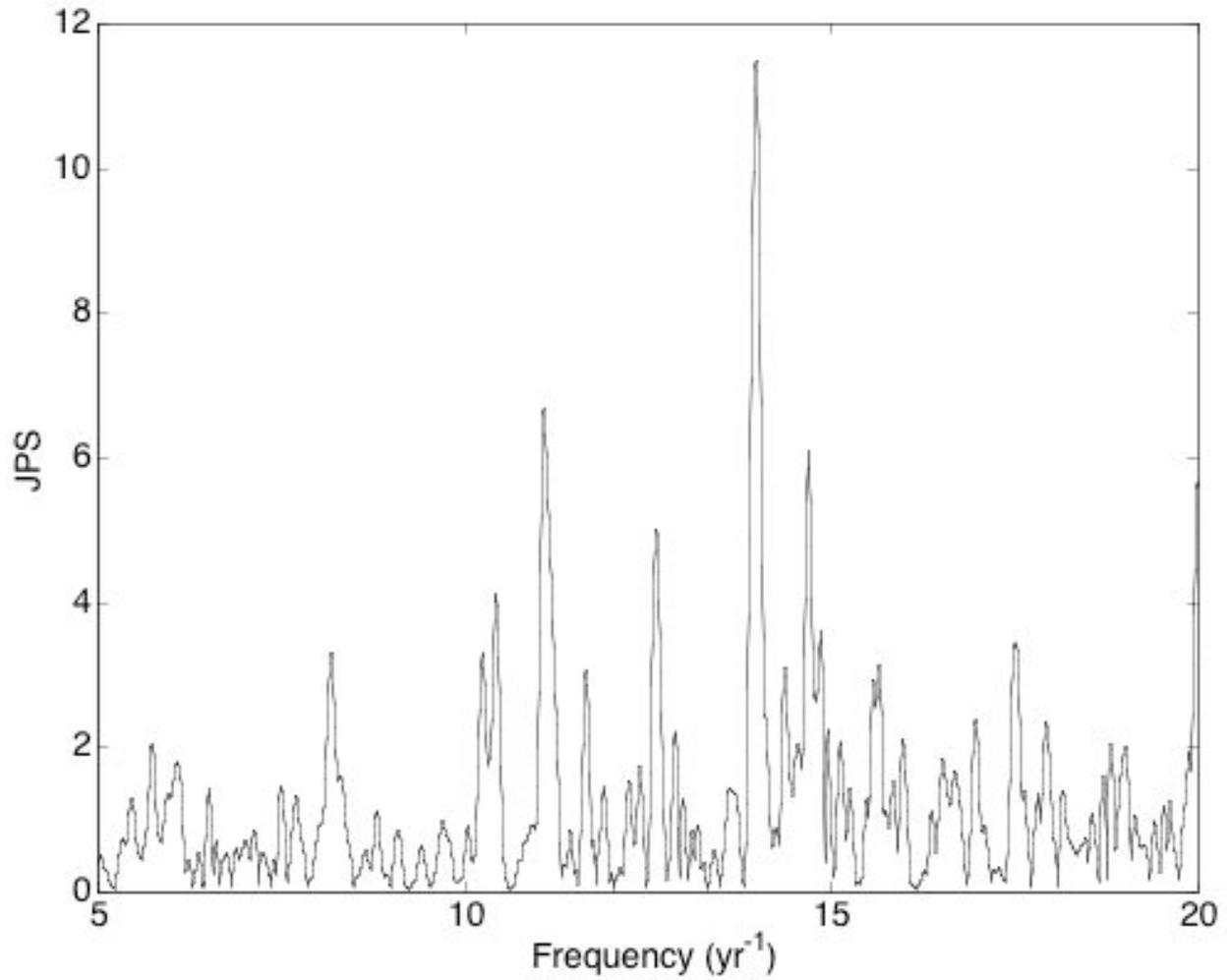

Figure 3. Joint Power Statistic formed from the m = 1, l = 2, 3, 4, 5, and 6 E-type r-mode frequencies in the Super-Kamiokande power spectrum.